\documentclass[pra,aps,twocolumn,nofootinbib,superscriptaddress,floatfix]{revtex4-1}
\usepackage{amsmath,amscd,amssymb,color}
\usepackage{graphicx,amsfonts,epsf}
\usepackage{epstopdf}
\usepackage{float}
\usepackage{enumerate}

\usepackage{color,hyperref}
\definecolor{darkblue}{rgb}{0.5,0.0,0.0}
\hypersetup{colorlinks,breaklinks,
linkcolor=darkblue,urlcolor=darkblue,
anchorcolor=darkblue,citecolor=
darkblue,pdfauthor=KavitaDorai, pdftitle=KavitaDoraiANN}

\begin{document}
\title{Neural network assisted quantum state and process
tomography using 
limited data sets}
\author{Akshay Gaikwad$^{\#}$}
\email{ph16010@iisermohali.ac.in}
\affiliation{Department of Physical Sciences, Indian
Institute of Science Education \& 
Research Mohali, Sector 81 SAS Nagar, 
Manauli PO 140306 Punjab India.}
\author{Omkar Bihani$^{\#}$}
\email{omkarbihani10@gmail.com}
\affiliation{Department of Physical Sciences, Indian
Institute of Science Education \& 
Research Mohali, Sector 81 SAS Nagar, 
Manauli PO 140306 Punjab India.}
\author{Arvind}
\email{arvind@iisermohali.ac.in}
\affiliation{Department of Physical Sciences, Indian
Institute of Science Education \& 
Research Mohali, Sector 81 SAS Nagar, 
Manauli PO 140306 Punjab India.}
\affiliation{Vice Chancellor, Punjabi University Patiala,
147002, Punjab, India}
\author{Kavita Dorai}
\email{kavita@iisermohali.ac.in}
\affiliation{Department of Physical Sciences, Indian
Institute of Science Education \& 
Research Mohali, Sector 81 SAS Nagar, 
Manauli PO 140306 Punjab India.}
\begin{abstract}
In this study we employ a feed-forward artificial neural network (FFNN)
architecture to perform tomography of quantum states and processes obtained from
noisy experimental data. To evaluate the performance of the FFNN, we use a
heavily reduced data set and show that the density and process matrices of
unknown quantum states and processes can be reconstructed with high fidelity.
We use the FFNN model to tomograph 100 two-qubit and 128 three-qubit states
which were experimentally generated on a nuclear magnetic resonance (NMR)
quantum processor.  The FFNN model is further used to characterize different
quantum processes including two-qubit entangling gates, a shaped pulsed field
gradient, intrinsic decoherence processes present in an NMR system, and various
two-qubit noise channels (correlated bit flip, correlated phase flip and a
combined bit and phase flip).  The results obtained via the FFNN model are
compared with standard quantum state and process tomography methods and the
computed fidelities demonstrates that for all cases, the FFNN model outperforms
the standard methods for tomography. 
\end{abstract} 
\maketitle
\def\thefootnote{$\#$}\footnotetext{These authors contributed equally to
this work}
\section{Introduction}
\label{sec1}
Quantum state tomography (QST) and quantum process tomography (QPT) are
essential techniques to characterize unknown quantum states and processes
respectively, and to evaluate the quality of quantum
devices~\cite{nielsen-book-10,childs-pra-2001,obrien-prl-04}.  Numerous
computationally and experimentally efficient QST and QPT algorithms have been
designed such as self-guided tomography\cite{chapman-prl-2016}, adaptive
tomography~\cite{pogo-pra-2017}, compressed sensing based QST and QPT protocols
which use heavily reduced data sets~\cite{riofrao-natcom-2017,gaikwad-qip-2021},
selective QPT~\cite{paz-prl-2008,gaikwad-pra-2018,gaikwad-sr-2022}, and direct
QST/QPT using weak measurements~\cite{kim-natcom-2018}.

Recently, machine learning (ML) techniques have been used to improve the
efficiency of tomography
protocols~\cite{carleo-science-2017,carleo-rmp-2019,torlai-arcm-2020}.  QST was
performed on entangled quantum states using a restricted Boltzmann machine based
artificial neural network (ANN) model~\cite{torlai-np-2018} and was
experimentally implemented on an optical system~\cite{neugebauer-pra-2020}.  ML
based adaptive QST was performed which adapts to 
experiments and
suggests suitable further measurements~\cite{quek-npj-2021}.  QST using
an attention based generative network was realized experimentally on an IBMQ
quantum computer\cite{cha-mlst-2021}.  ANN enhanced QST was carried
out after minimizing state preparation and measurement errors when
reconstructing the state on a photonic quantum dataset~\cite{palmieri-npj-2020}.
A convolutional ANN model was employed to reconstruct quantum states
with tomography measurements in the presence of simulated
noise~\cite{lohani-mlst-2020}.  Local measurement-based QST via ANN
was experimentally demonstrated on NMR\cite{xin-npj-2019}.  ML was used to
detect experimental multipartite entanglement structure for NMR entangled
states~\cite{advqtmtech-2022-ent}.  ANN was used to perform QST
while taking into account measurement imperfections~\cite{ijtp-2022-qst} and
were trained to uniquely reconstruct a quantum state without requiring any prior
information about the state~\cite{njp-2021-qst}.  ANN was used to
reconstruct quantum states encoded in the spatial degrees of freedom of photons
with high fidelity~\cite{npj-2020-qpt}.  ML methods were used to directly
estimate the fidelity of prepared quantum states~\cite{prl-2021-state-estimate}.
ANN was used to reconstruct quantum states in the presence of
various types of noise~\cite{pra-2022-nn-qst}.  Quantum state tomography in
intermediate-scale quantum devices was performed using conditional generative
adversial networks~\cite{prl-2021-gann}.

In this study, we employed a Feed Forward Neural Network
(FFNN) architecture to perform quantum state as well as
process tomography. We trained and tested the model on
states/processes generated computationally and then
validated it on noisy experimental data generated on an NMR
quantum processor.  Furthermore, we tested the efficacy of
the FFNN model on a heavily reduced data set, where a random
fraction of the total data set was used.  The FFNN model was
able to reconstruct the true quantum states and quantum
processes with high fidelity even with this heavily reduced
data set.

This paper is organized as follows: Section~\ref{sec2}
briefly describes the basic framework of the FFNN model in
the context of QST and QPT; Section~\ref{sec2.1} describes
the FFNN architecture while Section~\ref{sec2.2} details how
to construct the FFNN training data set to perform QST and
QPT.  Sections~\ref{sec3} and \ref{sec4} contain the results
of implementing the FFNN to perform QST and QPT of
experimental NMR data, respectively.  Section~\ref{sec5}
contains a few concluding remarks.
\section{FFNN Based QST and QPT}
\label{sec2}
\begin{figure}[t]
\centering
\includegraphics[angle=0,scale=1]{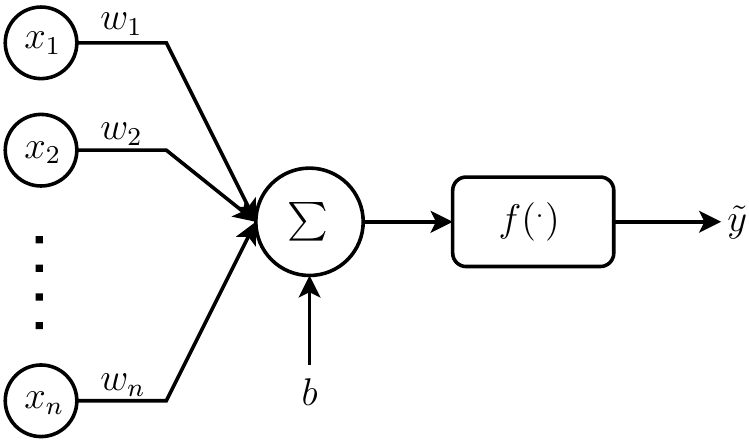}
\caption{(Color online) Basic unit of an ANN model,  
where the $x_i$ are the inputs, $w_i$ are the weights, 
$b$ is the bias, $\sum$ is the summation function, 
$f$ is an activation function and
$\tilde{y}$ is the output of the ANN.}
\label{nn0}
\end{figure}
\subsection{The Basic FFNN Architecture}
\label{sec2.1}
First we describe the multilayer perceptron model also referred to as a
Feed-Forward-Neural network (FFNN) which we employ to the task of characterizing
quantum states and processes. 
An ANN is a mathematical 
computing model motivated by the biological nervous system which
consists of adaptive units called neurons which are connected to other neurons
via weights.  A neuron is activated when its value is greater than a `threshold
value' termed the bias.  Figure~\ref{nn0} depicts a schematic of an ANN with $n$
inputs $x_1,x_2,\cdots,x_n$ which are connected to a neuron with weights
$w_1,w_2,\cdots,w_n$; the weighted sum of these inputs is compared with the
bias $b$ and is acted upon the activation function $f$, with the output
$\tilde{y}=f(\sum_{i=1}^{n} w_i x_i - b)$.

A multilayer 
FFNN architecture consists of three layers: the input layer, the
hidden layer and the output layer.  Data is fed into the input layer, which is
passed on to the hidden layers  and finally from the last hidden layer, it
arrives at the output layer.  Figure~\ref{flowchart} depicts a schematic of a
prototypical FFNN model with one input layer, two hidden layers and one output
layer,  which has been employed (as an illustration) to perform QST of an
experimental two-qubit NMR quantum state, using a heavily reduced data set.

The data is divided into two parts: a training dataset which is used to train
the model, a process in which network parameters, (weights and biases) are
updated based on the outcomes and a test dataset which is used to evaluate the
network {performance.  Consider `$m$' training elements
$\lbrace
(\vec{x}^{(1)},\vec{y}^{(1)}),(\vec{x}^{(2)},\vec{y}^{(2)}),\cdots
,(\vec{x}^{(p)},\vec{y}^{(p)}) \rbrace$ 
where $\vec{x}^{(i)}$ is the
$i^{th}$ input and 
$\vec{y}^{(i)}$ 
is the corresponding output.
Feeding these inputs to the network produces the outputs
$[\tilde{\vec{y}}^{(1)},\tilde{\vec{y}}^{(2)},...,\tilde{\vec{y}}^{(p)}]$.
Since network parameters are initialized randomly, the predicted output is not
equal to the expected output. 
Training of this network can be achieved by minimizing
a mean-squared-error 
cost function, 
with respect to the network
parameters, by using a stochastic gradient 
descent method and the backpropagation
algorithm~\cite{gradient-descent}: 
\begin{align} w_{ij}
\rightarrow w'_{ij} =& w_{ij}-\frac{\eta}{p'} \sum_{i=1}^{p'}
\frac{\partial}{\partial  w_{ij}} \mathcal{L}(\vec{x}^{(i)}) \\ b_{i}
\rightarrow b'_{i} =& b_{i}-\frac{\eta}{p'} \sum_{i=1}^{p'}
\frac{\partial}{\partial  b_{i}} \mathcal{L}(\vec{x}^{(i)}) 
\label{gradient-eqn}
\end{align} 
where 
$\mathcal{L}(x^{(i)})= ||\vec{y}^{(i)}-
\tilde{\vec{y}}^{(i)}||^2$
is the cost function of the randomly chosen $m'$ training inputs $x^{(i)}$,
$\eta$ is the learning rate and $w'_{ij}$ and $b'_{i}$ are updated weights and
biases, respectively.

\begin{figure*}[ht]
\centering
\includegraphics[angle=0,scale=1]{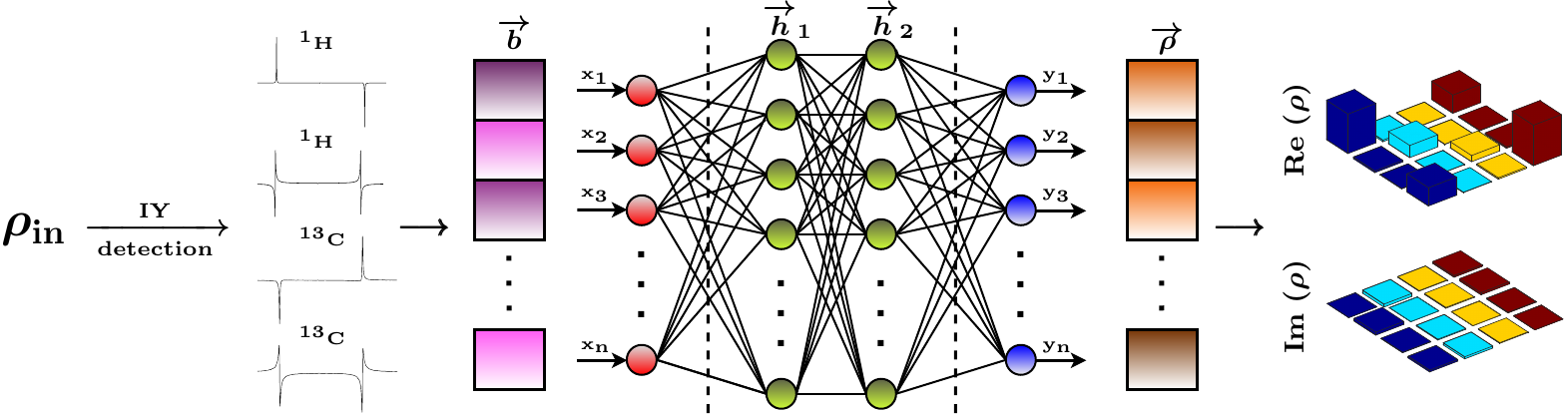} 
\caption{(Color online) Flowchart illustrating  the FFNN model used to perform
QST on two-qubit quantum states generated on an NMR quantum; on the left,
$\rho_{in}$ represents the state which is to be tomographed; ${\rm IY}$ denotes
a tomographic operation, which is followed by signal detection, the set of
depicted NMR spectra are those obtained after the tomographic measurement.  The
FFNN with two hidden layers is represented next, which then uses a reduced data
set to reconstruct the final experimental tomographs represented on the right.}
\label{flowchart}
\end{figure*}
\subsection{FFNN Training Dataset for QST and QPT}
\label{sec2.2}
An $n$-qubit density operator $\rho$
can be expressed as a matrix in 
the product basis by:
\begin{equation} 
\rho =
\sum_{i=0}^{3}\sum_{j=0}^{3}...\sum_{n=0}^{3}
a_{ij...n}\sigma_i \otimes \sigma_j \otimes...\sigma_n
\label{e3} 
\end{equation} 
where $a_{00...0} = 1/2^n $, $\sigma_0$ denotes the $2 \times 2$ identity matrix
and $\sigma_i, i=1,2,3$ are single-qubit Pauli
matrices. 

The aim of QST is to reconstruct $\rho$ from a set of tomographic measurements.
The standard procedure for QST involves solving linear system 
of equations of the form~\cite{long-job-2001}:
\begin{equation} 
\mathcal{A} \mathcal{X} = \mathcal{B}
\label{e4}
\end{equation}
where $\mathcal{A}$ is a fixed coefficient matrix and only depends on the chosen
measurement settings, $\mathcal{X}$ is a
column matrix which contains elements of the density matrix which needs to be
reconstructed, and the input vector $\mathcal{B}$
contains the actual experimental data. 

The FFNN model is trained on a dataset containing randomly
generated pure and mixed states.  To generate these
ensembles, consider a normal distribution
$\mathcal{N}(\mu=0,\sigma^2=1)$ with zero mean and unit
variance.
An $n$-qubit pure random state in the computational basis is
represented by an $2^n$ column vector $C$ whose $i$th entry
$c_i$ generated from the random distribution as follows: 
\begin{equation} 
\label{e7}
c_i  = \frac{1}{N}(\mathfrak{d}_{i}+ i\, \mathfrak{e}_{i})
\end{equation} 
where $\mathfrak{d}_{i},\mathfrak{e}_{i}$ 
are randomly chosen from the distribution $\mathcal{N}$ and
$N$ is a normalization factor to ensure that $C$ represent a
unit vector.

For mixed states:
\begin{equation} 
\label{e8}
R = \mathcal{D}_{i} + i\, \mathcal{E}_{i}
\end{equation}  
where $R$ is a $2^n \times 2^n$ matrix with its elements 
$\mathcal{D}_{i},\mathcal{E}_{i}$
randomly sampled 
from the
normal distribution $\mathcal{N}$.
Using the $R$ matrix, the
corresponding mixed state density matrix $\rho_{\rm{mix}}$ is constructed as
$\rho_{\rm{mix}} = \frac{R R^{\dagger}}{\rm{Tr}(R R^{\dagger})}$.

The FFNN is trained 
on both pure as well as mixed states and the
appropriate density matrices are generated.
After generating the density matrices 
$\mathcal{X}_i$, 
the corresponding $\mathcal{B}_i$ are
computed using Eq.(\ref{e4}).  The training elements 
$\lbrace \mathcal{B}_{i},\mathcal{X}_{i} \rbrace$ 
are then used to train the FFNN model given in
Figure~\ref{flowchart}, where 
$\mathcal{B}_{i}$ 
are the inputs to the FFNN and
$\mathcal{X}_{i}$
are the 
corresponding labeled outputs.

QPT of given a quantum process is typically performed
using the Kraus operator representation, wherein for a
fixed operator basis set $\lbrace E_i \rbrace$, a quantum map $\Lambda$ acting
on an input state $\rho_{{\rm in}}$ can be written as~\cite{kraus-book-1983}:
\begin{equation}
\Lambda(\rho_{in}) = \sum_{m,n} \chi_{mn} E_m \rho_{in} E_n^{\dagger}
\label{e5}
\end{equation}
where $\chi_{mn}$ are
the elements of the process matrix $\chi$ characterizing the quantum map $\Lambda$. 
The $\chi$ matrix can be experimentally determined by
preparing a complete set of linearly independent input states,
estimating the output states after action of the map,  and
finally computing the elements of $\chi_{mn}$ 
from these experimentally estimated output states via
linear equations of the form\cite{chuang-jmo-09}: 
\begin{equation}
\beta \vec{\chi} = \vec{\lambda}
\label{e6}
\end{equation}
where $\beta$ is a coefficient matrix,
$\vec{\chi}$ contains the elements $\lbrace \chi_{mn} \rbrace$ which are to be
determined and $\vec{\lambda}$ is a vector representing 
the experimental data.

The training data set for using the FFNN model to perform QPT is constructed by
randomly generating a set of unitary operators.  The generated unitary operators
are allowed to act upon the input states $ \rho_{in} = \lbrace |0\rangle,
|1\rangle, \frac{1}{\sqrt{2}}(\vert 0 \rangle + \vert 1
\rangle),\frac{1}{\sqrt{2}}(\vert 0
\rangle + i\vert 1 \rangle) \rbrace^{\otimes n}$, to obtain
$\rho_{out} = U \rho_{in} U^{\dagger}$.  All the output
states $\rho_{out}$ are
stacked to form form $\vec{\lambda}$.  Finally, $\vec{\chi}$ is computed using
Eq.~(\ref{e6}).  The training elements $\lbrace \vec{\lambda}_{i},
\vec{\chi}_{i} \rbrace$ will then be used to train the FFNN model, where
$\vec{\lambda}_{i}$ acts as the input to FFNN and $\vec{\chi}_{i}$ is the
corresponding labeled output.

\begin{table} 
\centering 
\caption{Average state fidelities obtained after training the FFNN model to
perform QST on 3000 test two-qubit states ($M_{{\rm data}} = 20$) for training
datasets of different sizes, with the number of epochs varying from 50 to 150
for each dataset 
(an epoch refers to one iteration of the complete training dataset). 
}
\setlength{\tabcolsep}{8pt} 
\renewcommand{\arraystretch}{1.1} 
\begin{tabular}{c | c | c | c}
\hline Dataset Size & \multicolumn{3}{c}{Fidelity} \\
\hline 
& Epoch(50) & Epoch(100) & Epoch(150)\\
\hline 500 & 0.8290 & 0.9176 & 0.9224 \\ 1000 & 0.9244 & 0.9287 & 0.9298 \\ 5000
& 0.9344 & 0.9379 & 0.9389 \\ 10000 & 0.9378 & 0.9390 & 0.9400 \\ 20000 & 0.9394
& 0.9414 & 0.9409 \\ 80000 & 0.9426 & 0.9429 & 0.9422 \\ \hline \end{tabular}
\label{2qst_epoch} \end{table}
\begin{table} 
\centering 
\caption{Average state fidelities obtained after training the FFNN model to
perform QST on 3000 test three-qubit states ($M_{{\rm data}} = 120$) for
training datasets of different sizes, with the number of epochs varying from 50
to 150 for each dataset
(an epoch refers to one iteration of the complete training dataset).}
\setlength{\tabcolsep}{8pt} 
\renewcommand{\arraystretch}{1.1} 
\begin{tabular}{c | c | c | c}
\hline Dataset Size & \multicolumn{3}{c}{Fidelity} \\ 
\hline
& Epoch(50) & Epoch(100) & Epoch(150)\\
\hline 500 & 0.6944 & 0.8507 & 0.8716 \\ 1000 & 0.8793 & 0.8994 & 0.9025 \\ 5000
& 0.9231 & 0.9262 & 0.9285 \\ 10000 & 0.9278 & 0.9321 & 0.9332 \\ 20000 & 0.9333
& 0.9362 & 0.9393 \\ 80000 & 0.9413 & 0.9433 & 0.9432 \\ \hline \end{tabular}
\label{3qst_epoch} \end{table}
\begin{table} 
\centering 
\caption{Average process fidelities obtained after training the FFNN model to
perform QPT on 3000 test three-qubit states ($M_{{\rm data}} = 200$) for
training datasets of different sizes, with the number of epochs varying from 50
to 150 for each dataset
(an epoch refers to one iteration of the complete training dataset).}
\setlength{\tabcolsep}{8pt} 
\renewcommand{\arraystretch}{1.1} 
\begin{tabular}{c | c | c | c}
\hline Dataset Size & \multicolumn{3}{c}{Fidelity} \\ 
\hline
& Epoch(50) & Epoch(100) & Epoch(150)\\
\hline 500 & 0.4904 & 0.5421 & 0.5512 \\ 2000 & 0.6872 & 0.7090 & 0.7202 \\
15000 & 0.7947 & 0.8128 & 0.8218 \\ 20000 & 0.8047 & 0.8203 & 0.8295 \\ 50000 &
0.8305 & 0.8482 & 0.8598 \\ 80000 & 0.8441 & 0.8617 & 0.8691 \\ \hline
\end{tabular} \label{2qpt_epoch} \end{table}

To perform tomography of given state or process one has to perform a series
of experiments.  Then the input vector is constructed where the entries are
outputs/readouts of the experiments. In the case of standard QST or QPT, a
tomographically complete set of experiments needs to be done and the input
vector corresponding to the tomographically complete set of experiments is
referred to as the full data set.  
To perform QST and QPT via FFNN on a heavily reduced data set of size $m$, a
reduced size input vector $\vec{b}_{m}$ with fewer elements (and a
correspondingly reduced $\vec{\lambda}_{m}$) is constructed by randomly
selecting $m$ elements from the input vectors while the remaining elements are
set to 0 (zero padding); these reduced input vectors together with the
corresponding labeled output vectors are used to train the FFNN.

The FFNN was trained and implemented using the Keras
Python library~\cite{keras} with the Tensor-Flow backend, on an Intel Xeon
processor with 48GB RAM and a CPU base speed of 3.90GHz.
To perform QST and QPT the LeakyReLU ($\alpha=0.5$) activation function was used
for both the input and the hidden layers of the FFNN:
\begin{equation} \begin{split}
\text{LeakyReLU(x)} = & x\, ; \, x>0 \\ = & \alpha x\, ; \, x<0 \end{split}
\end{equation} 
A linear activation function was used for the output layer.  A cosine similarity
loss function, $\mathcal{L} = \arccos \left( \frac{\hat{\vec{y}} .
\tilde{\vec{y}}}{ ||\hat{\vec{y}}||.||\tilde{\vec{y}}|| } \right)$ was used for
validation and the {\em adagrad} ($\eta=0.5$) optimizer with a learning rate
$\eta$, was used to train the network.  The {\em adagrad} optimizer adapts the
learning rate relative to how frequently a parameter gets updated during
training.  

The FFNN was used to perform QST on 3000 two-qubit and three-qubit test quantum
states and to perform QPT on 3000 two-qubit test quantum processes for training
datasets of different sizes. 
The number of epochs were varied from 50 to 150 for each dataset, where
an epoch refers to one iteration of the training dataset during the
FFNN training process.  
The computed average fidelities of 3000 two-qubit and three-qubit
test quantum states and 3000 two-qubit test quantum processes are shown in
Tables~\ref{2qst_epoch}, \ref{3qst_epoch} and \ref{2qpt_epoch}, respectively;
$M_{{\rm data}}$ refers to the reduced size of the data set.
After comparing the effect of training data size, the value of
$M_{{\rm data}}$ and the number of epochs, 
the maximum size of the training data set was chosen  to be 80000 and the
maximum number of epochs was set to 150 for performing QST and QPT.
After 150 training epochs the validation loss function remained constant.

\section{FFNN Based QST on Experimental Data} 
\label{sec3}
We used an NMR quantum processor as the experimental testbed to generate data
for the FFNN model.  We applied FFNN to perform QST of two-qubit and three-qubit
quantum states using a heavily
reduced data set of noisy data generated on an NMR quantum processor.  The
performance of the FFNN was evaluated by computing the average state or average
process fidelity. The state fidelity is given by~\cite{suter-prl-2014}:
\begin{equation}
\label{fid}
\mathcal{F}
=\frac{\left|\operatorname{Tr}\left[\rho_{{\rm FFNN}} 
\rho_{{\rm STD }}^{\dagger}\right]\right|}
{\sqrt{\operatorname{Tr}
\left[\rho_{{\rm FFNN }}^{\dagger} 
\rho_{{\rm FFNN }}\right] 
\operatorname{Tr}\left
[\rho_{{\rm STD }}^{\dagger} \rho_{{\rm STD }}\right]}}
\end{equation}
where $\rho_{{\rm FFNN }}$ and $\rho_{{\rm STD }}$ are the density matrices
obtained via the FFNN and the standard linear inversion method, respectively.
The process fidelity can be computed by replacing the $\rho$ in Eq.~(\ref{fid})
by $\chi$, where $\chi_{{\rm FFNN }}$ and $\chi_{{\rm STD }}$ are process
matrices obtained via the FFNN and standard linear inversion method,
respectively.
\begin{figure}
\centering
\includegraphics[angle=0,scale=1.0]{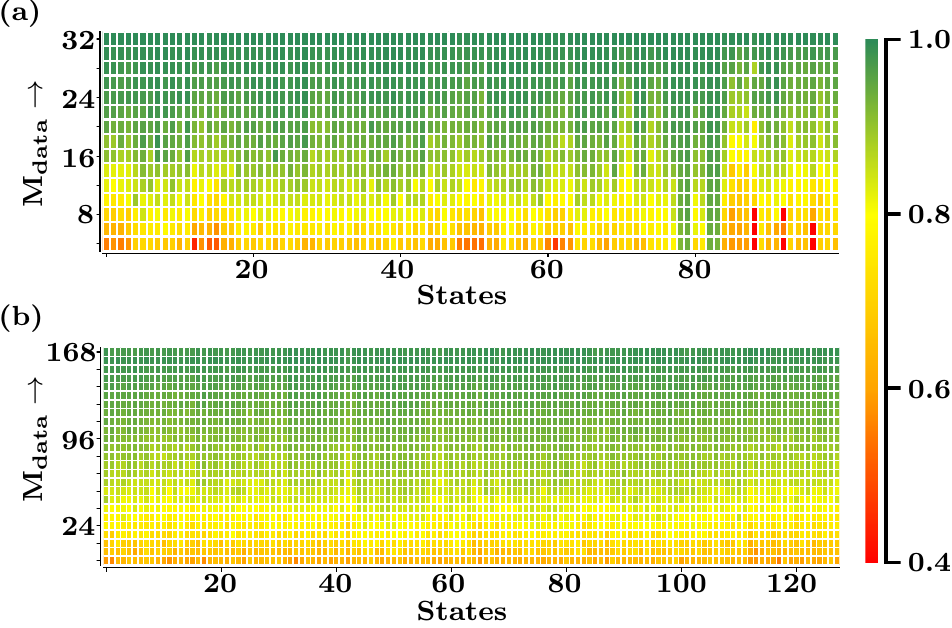} 
\caption{(Color online) Fidelity ($\bar{\mathcal{F}}$) between the FFNN model
and the standard linear inversion method vs size of the 
heavily reduced dataset ($M_{data}$),
for QST performed on (a) 100 two-qubit states and (b) 128 three-qubit states
respectively.  The states are numbered on the $x$-axis and the color coded bar
on the right represents the value of the fidelity.} 
\label{qst}
\end{figure}
\begin{figure}
\centering
\includegraphics[angle=0,scale=0.88]{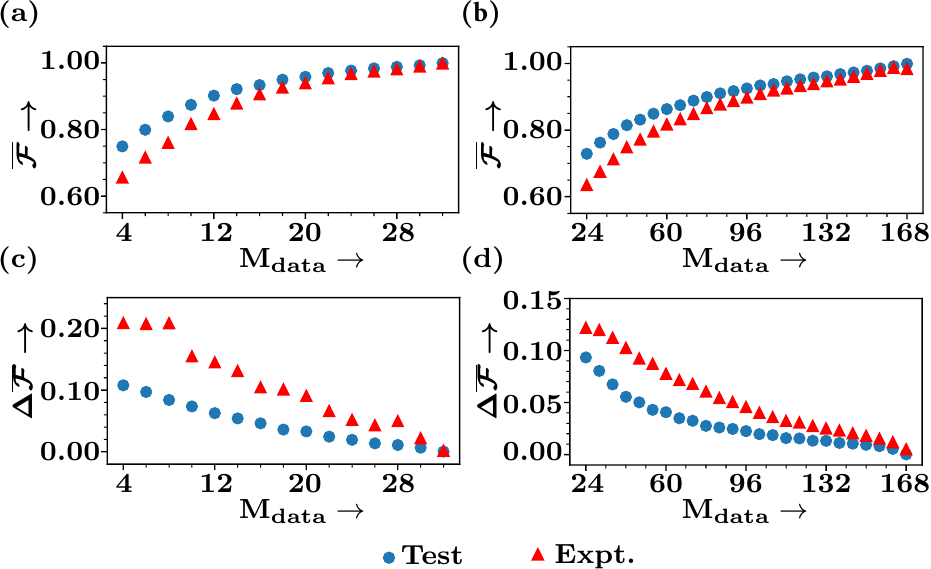} 
\caption{(Color online) Average fidelity ($\bar{\mathcal{F}}$) and standard
deviation $\Delta \bar{\mathcal{F}}$ plotted as a function of the size of the
heavily reduced 
dataset ($M_{data}$) computed for FFNN based QST on two qubits ((a) and (c)) and
on three qubits ((b) and (d)), respectively.  The average fidelity for the test
dataset (blue dots) is calculated for 3000 states while for experimental
data-set the average fidelity is calculated by randomly choosing
the reduced dataset $M_{data}$ elements from the full set
for 100 2-qubit states and 128 3-qubit states, and then repeating the
procedure 50 times.
}
\label{avgqst}
\end{figure}
\begin{figure}
\includegraphics[angle=0,scale=1]{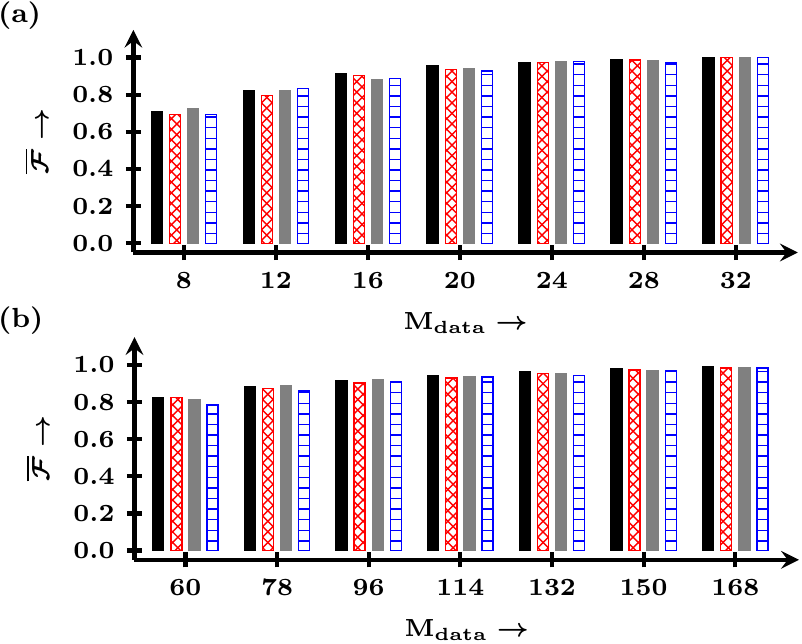} 
\caption{(Color online) Fidelity ($\bar{\mathcal{F}}$) versus size of the
heavily reduced dataset ($M_{{\rm data}}$) computed for FFNN based QST of (a)
two-qubit Bell states, where 
the different bars correspond to
four different Bell states, and (b) 
three-qubit GHZ (black and red cross-hatched bars) 
and Biseparable states (gray and horizontal blue bars).
}
\label{bell_ghz}
\end{figure}

QST of a two-qubit NMR system is typically performed using a set of four unitary
rotations: $\lbrace II, IX, IY, XX \rbrace$ where $I$ denotes the identity
operation and $X(Y)$ denotes a $90^{\circ}$ $x$ rotation on the specified qubit.
The input vector $\vec{b}$ (Eq.~(\ref{e4}))  is constructed by applying the
tomographic pulses followed by measurement, wherein the signal which is recorded
in the time domain is then Fourier transformed to obtain the NMR spectrum. For
two qubits, there are four peaks in the NMR spectrum and each measurement yields
eight elements of the vector $\vec{b}$;  the dimension of the input vector
$\vec{b}$ is $33 \times 1$ (32 from tomographic pulses and 1 from the unit trace
condition).  Similarly, QST of a three-qubit NMR system is typically performed
using a set of seven unitary rotations: $\lbrace III, IIY, IYY, YII, XYX, XXY,
XXX \rbrace$.  Each measurement produces 12 resonance peaks in the NMR spectrum
(4 per qubit); the dimension of the input vector $\vec{b}$ is be $169 \times 1$.
To evaluate the performance of FFNN model in achieving full QST of two-qubit and
three-qubit states, we experimentally prepared 100 two-qubit states and 128
three-qubit states using different preparation settings and calculated the
average fidelity between the density matrix predicted via the FFNN model and
that obtained using the standard linear inversion method for QST.  We also
performed full FFNN based QST of maximally entangled two-qubit Bell states and
three-qubit GHZ and Biseparable states using a heavily reduced data set.

The FFNN model was trained on 80,000 states to perform QST.  To perform FFNN
based QST on two- and three-qubit states, we used three hidden layers containing
100, 100 and 50 neurons and 300, 200 and 100 neurons, respectively.  The
performance of the trained FFNN is shown in Figure~\ref{qst}.  The fidelity
between density matrices obtained via FFNN and standard linear inversion method
of 100 experimentally generated two-qubit states and of 128 experimentally
generated three-qubit states  is shown in Figures~\ref{qst}(a) and (b),
respectively.  The reduced input vector of size $M_{{\rm data}}$ is plotted on
the $y$-axis and the quantum states are numbered along the $x$-axis.

The performance of the FFNN for QST is evaluated in Figure~\ref{avgqst}  by
computing the average state fidelity $\mathcal{\bar{F}}$ calculated over a set
of test/experimental states.  The reduced size $M_{data}$ of the input vector
which was fed into the FFNN is plotted along the $x$-axis.  The average fidelity
$\mathcal{\bar{F}}$ and the standard deviation $\sigma$ in average state
fidelity $\mathcal{\bar{F}}$  are plotted along the $y$-axis in (a) and (c) for
two-qubit states and in (b) and (d) for three-qubit states, respectively.  For a
given value of $M_{data}$, the average fidelity
$\mathcal{\bar{F}}_i=\frac{1}{50}\sum_{n=1}^{50} \mathcal{F}_n$ of a given
quantum state $\rho_i$ predicted via FFNN is calculated by randomly selecting
$M_{data}$ elements from the corresponding full input vector $\vec{b}$ for 50
times.  For test data sets (blue circles), the performance of the FFNN is
evaluated by computing the average fidelity $\mathcal{\bar{F}} =
\frac{1}{3000}\sum_{n=1}^{3000} \mathcal{\bar{F}}_n$ over 3000 two-qubit and
three-qubit states.  For experimental data sets (red triangles), the performance
of the FFNN is evaluated by computing the average fidelity $\mathcal{\bar{F}}$
over 100 two-qubit and 128 three-qubit states, respectively.  
The standard deviation $\sigma$ in average
state fidelity $\mathcal{\bar{F}}$ is: 
\begin{equation}
\sigma = \sqrt{\frac{\sum_{i=1}^{N}
(\mathcal{\bar{F}}_i-\bar{\mathcal{F}})^2}{N-1}}
\label{e10}
\end{equation}
As inferred from Figure~\ref{avgqst}, the FFNN model is able to predict  an
unknown two-qubit test state with average fidelity $\mathcal{\bar{F}} \geq
0.8392 \pm 0.084$ for a reduced data set of size $M_{data} \geq 8$, and is able
to predict an unknown three-qubit test state with average fidelity
$\mathcal{\bar{F}} \geq 0.8630 \pm 0.0407$ for a reduced data set of  size
$M_{data} \geq 60$.  Similarly, for experimental quantum states, the FFNN model
is able to predict two-qubit states with an average fidelity $\mathcal{\bar{F}}
\geq 0.8466 \pm 0.1450$ for a reduced data set of size $M_{data} \geq 12$, while
for three-qubit experimental states, the FFNN is able to predict the unknown
quantum state with average fidelity $\mathcal{\bar{F}} \geq 0.8327 \pm 0.0716$
for a reduced data set of size $M_{data} \geq 60$.  
When the full input vector
$\vec{b}$ is considered, the average fidelity calculated over 3000 two- and
three-qubit test states turns out to be $\mathcal{\bar{F}} = 0.9993 $ and
$\mathcal{\bar{F}} = 0.9989 $, respectively. The average fidelity calculated
over 100 two-qubit and 128 three-qubit experimental states turns out to be
$\mathcal{\bar{F}} = 0.9983 $ and $\mathcal{\bar{F}} = 0.9833 $, respectively,
for the full input data set.

The FFNN model was applied to perform QST of two-qubit maximally entangled Bell
states and three-qubit GHZ and biseparable states.  Figure~\ref{bell_ghz}
depicts the experimental fidelities $\mathcal{F(\rho_{\rm \small FFNN},\rho_{\rm
\small STD})}$ of two-qubit Bell states and three-qubit GHZ and biseparable
states calculated between the density matrices predicted via FFNN and those
obtained via standard linear inversion QST for a reduced data set of size
$M_{\rm data}$.  The black, crosshatched red, gray and horizontal blue bars in
Figure~\ref{bell_ghz}(a) correspond to the Bell states $\vert B_1 \rangle =
(\vert 00 \rangle + \vert 11 \rangle)/\sqrt{2} $, $\vert B_2 \rangle = (\vert 01
\rangle - \vert 10 \rangle)/\sqrt{2} $, $\vert B_3 \rangle = (\vert 00 \rangle -
\vert 11 \rangle)/\sqrt{2} $ and $\vert B_4 \rangle = (\vert 01 \rangle + \vert
10 \rangle)/\sqrt{2} $, respectively.  The black and red cross-hatched bars in
Figure~\ref{bell_ghz}(b) correspond to three-qubit GHZ states $\vert \psi_1
\rangle = (\vert 000 \rangle + \vert 111 \rangle)/\sqrt{2}$ and $\vert \psi_2
\rangle = (\vert 010 \rangle + \vert 101 \rangle)/\sqrt{2}$ respectively, while
the gray and horizontal blue bars correspond to three-qubit biseparable states
$\vert \psi_3 \rangle = (\vert 000 \rangle + \vert 001 \rangle + \vert 110
\rangle + \vert 111 \rangle)/2 $ and $\vert \psi_4 \rangle = (\vert 000 \rangle
+ \vert 010 \rangle + \vert 101 \rangle + \vert 111 \rangle)/2 $, respectively.
The bar plots in Figure~\ref{bell_ghz} clearly demonstrate that the FFNN model
is able to predict the two- and three-qubit entangled states with very high
fidelity for a reduced data set.

We note here in passing that the size of the heavily reduced dataset
$M_{\rm data}$ is equivalent to the number of experimental readouts which are
used to perform QST (QPT), while the standard QST (QPT) methods based on linear
inversion always use the full dataset. Hence, the highest value of $M_{\rm
data}$ is the same as the size of the full dataset which is 32, 168 and 256 for
two-qubit and three-qubit QST and for two-qubit QPT, respectively.
\section{FFNN Based QPT on Experimental Data} 
\label{sec4}
We used the FFNN model to perform two-qubit QPT for three different experimental
NMR data sets: (i) unitary quantum gates (ii) non-unitary processes such as
natural NMR decoherence processes and pulsed field gradient and (iii)
experimentally simulated correlated bit flip, correlated phase flip and
correlated bit+phase flip noise channels using the duality algorithm on an NMR
quantum processor.  
\subsection{FFNN Reconstruction of Two-Qubit Unitary and Non-Unitary Processes} 
\label{sec4.1}
The FFNN model was trained on 80,000 synthesized two-qubit quantum processes
using a heavily reduced data set, with three hidden layers containing 600, 400
and 300 neurons, respectively. The performance of the trained FFNN was evaluated
using 3000 test and 10 experimentally implemented quantum processes on NMR. 

\begin{figure}[t]
\centering
\includegraphics[angle=0,scale=0.9]{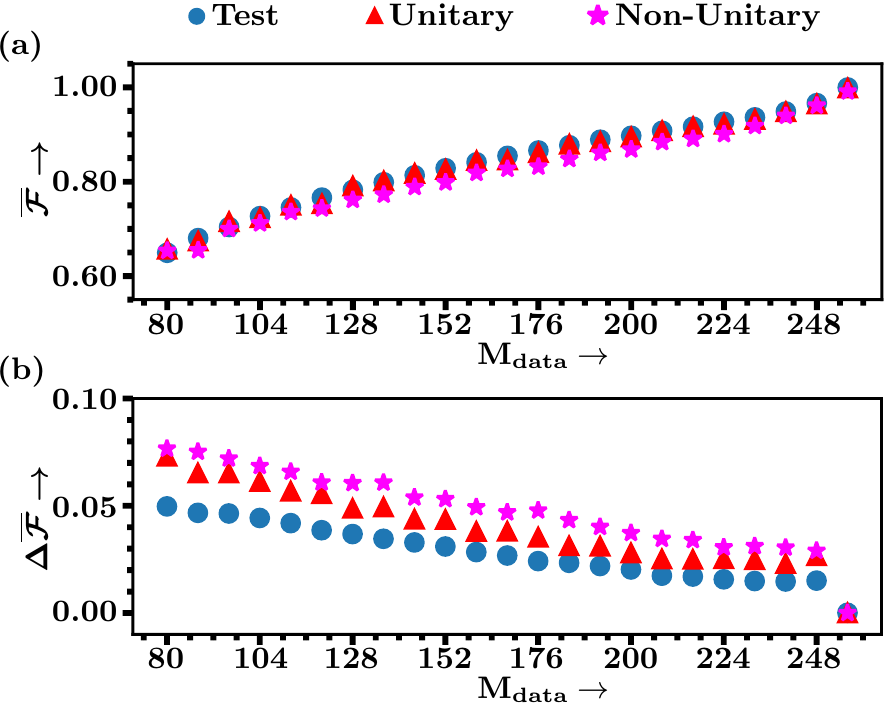} 
\caption{(Color online) (a) Average process fidelity ($\bar{\mathcal{F}}$) and
(b) Standard deviation $\Delta \bar{\mathcal{F}}$ obtained for QPT of two-qubit
processes using the FFNN model versus size of the dataset ($M_{data}$).  For the
test dataset (blue dots) the average fidelity is calculated for 3000 processes
while for experimental unitary processes (red triangles) and non-unitary
processes (magenta stars) the average fidelity is calculated by
randomly choosing the reduced dataset $M_{data}$ elements 
from the full set for four unitary
quantum gates and six non-unitary processes, 
and then repeating the procedure 300 times.
} 
\label{avgqpt}
\end{figure}

The FFNN results for QPT of various two-qubit experimental quantum processes are
shown in Figure~\ref{avgqpt}. The quality of the FFNN is evaluated by means of
the average process fidelity $\mathcal{\bar{F}}$, between the process matrix
predicted by the FFNN ($\chi_{\rm \small FFNN}$) using a reduced data set of
size $M_{data}$ and the process matrix obtained via the standard QPT method
($\chi_{\rm \small STD}$)  using a full data set.

Figure~\ref{avgqpt}(a) depicts the performance of the FFNN evaluated on 3000
two-qubit test quantum processes (blue circles), where the $y$-axis denotes the
average fidelity $\mathcal{\bar{F}} = \frac{1}{3000} \sum_{n=1}^{3000}
\mathcal{\bar{F}}_n$, where $\mathcal{\bar{F}}_n = \frac{1}{300}
\sum_{i=1}^{300} \mathcal{\bar{F}}_i $ is the average fidelity of the $n$th test
quantum process calculated by randomly constructing an input vector of given
size and repeating the process 300 times.  Similarly, the red triangles and pink
stars correspond to four unitary and six non-unitary quantum processes
respectively, obtained from experimental data.  The plots given in
Figure~\ref{avgqpt}(a) clearly show that the FFNN model is able to predict
unitary as well as non-unitary quantum processes from a noisy experimental
reduced data set, with good accuracy.  For instance, for $M_{{\rm data}} = 160$,
the FFNN is able to predict the test process with $ \mathcal{\bar{F}} = 0.8411
\pm 0.0284 $, whereas the experimental unitary and non-unitary processes are
obtained with $ \mathcal{\bar{F}}= 0.8447 \pm 0.038$ and $0.8187 \pm 0.0493$,
respectively.  Hence, the value of $M_{data}$ can be set accordingly, depending
on the desired accuracy and precision.  The standard deviation in average
fidelity $\mathcal{\bar{F}}$ is calculated using Eq.~(\ref{e10}) over 3000
quantum processes  and is depicted in Figure~\ref{avgqpt}(b).  From
Figure~\ref{avgqpt}(b), it can be observed that the FFNN model performs better
for the QPT of unitary processes as compared to non-unitary processes, since the
corresponding process matrices are more sparse.

The experimental fidelity obtained via FFNN of individual quantum processes is
given in Figure~\ref{qpt}, where the average fidelity is calculated for a set of
quantum processes for a given value of the reduced dataset $M_{data}$.  For the
test dataset the $ \mathcal{\bar{F}} $ is calculated over 3000 test processes,
whereas for the experimental data set, $ \mathcal{\bar{F}} $ is computed over
four unitary  and six non-unitary processes.  For  the unitary quantum gates:
Identity, CX180, CNOT and CY90, (corresponding to a `no operation' gate, a bit
flip gate, a controlled rotation about the $x$-axis by $180^{\circ}$ and a
controlled rotation about the $y$-axis by $90^{\circ}$, respectively) the FFNN
is able to predict the corresponding process matrix with average fidelities of
$\mathcal{\bar{F}} = 0.8767 \pm 0.0356, 0.8216 \pm 0.0463, 0.8314 \pm 0.0387 $
and $0.8489 \pm 0.0315$ respectively, using a reduced data set of size 160.  The
six non-unitary processes to be tomographed include free evolution processes for
two different times: $D1 = 0.05$ sec and $D2 = 0.5$ sec, a magnetic field
gradient pulse (MFGP), and three error channels, namely, a correlated bit flip
(CBF) channel, a correlated phase flip (CPF) channel, and a correlated bit-phase
flip  (CBPF) channel.  There are several noise channels acting simultaneously
all the qubits, during the free evolution times $D1$ and $D2$, such as the phase
damping channel (corresponding to the T$_2$ NMR relaxation process) and the
amplitude damping channel (corresponding to the T$_1$ NMR relaxation process).
The MFGP process is typically implemented using gradient coils in NMR hardware
where the magnetic field gradient is along the $z$-axis.  The MFGP process to be
tomographed is a  sine-shaped pulse of duration of 1000$\mu$s, 100 time
intervals = 100 and an applied gradient strength of 15\%.  For the intrinsic
non-unitary quantum processes D1 D2, and the MFGP, the FFNN is able to predict
the corresponding process matrix with average fidelities of $\mathcal{\bar{F}} =
0.8373 \pm 0.0381, 0.7607 \pm 0.0690$ and $0.7858 \pm 0.0703$ respectively,
using a reduced data set of size 160. It is evident from the computed fidelity
values that the FFNN performs better if the process matrix is sparse.
 
\begin{table} 
\centering \caption 
{\label{qptfid} Experimental fidelities
$\mathcal{F}$ computed between $\chi_{\small{\rm FFNN}}$, the process matrix
predicted via FFNN using a full data set, and $\chi_{\small{\rm STD}}$, the
process matrix obtained via the standard QPT method.  
}
\setlength{\tabcolsep}{3.5pt} 
\renewcommand{\arraystretch}{1.4} 
\begin{tabular}{|p{1.5cm} |p{1.5cm} ||p{2cm}|
p{1.5cm}|}
\hline \hline 
Unitary Process & ~~~$\mathcal{F}$& Non-Unitary
Process & ~~$\mathcal{F}$~~~~\\ 
\hline 
Test  & 0.9997 & D1 & 0.9987 \\ Identity
& 0.9943  & D2 & 0.9635 \\ CNOT & 0.9996 & Grad & 0.9917\\ CX180 & 0.9996 & CBF
& 0.9943\\ CY90 & 0.9996 & CPF & 0.9996\\ & & CBPF & 0.9996\\ 
\hline
\end{tabular} 
\end{table}

Although our main goal is to prove that the FFNN is able to
reconstruct quantum states and processes with a high fidelity even for
heavily reduced datasets, we also wanted to verify the efficacy of the
network when applied to a complete data set. 
The values of process fidelity obtained via FFNN for the full
data set are shown in Table~\ref{qptfid}, where it is clearly evident that the
FFNN is able to predict the underlying quantum process with very high fidelity,
and works accurately even for non-unitary quantum processes.  The somewhat lower
fidelity of the D2 process as compared to other quantum processes can be
attributed to the corresponding process matrix being less sparse.

\subsection{FFNN Reconstruction of Correlated Noise Channels}
\label{sec4.2}
The duality simulation algorithm (DSA) can be used to simulate fully correlated
two-qubit noise channels, namely the CBF, CPF and CBPF
channels~\cite{xin-pra-2017}.  The FFNN model is then employed to fully
characterize these channels. DSA allows us to simulate the arbitrary dynamics of
an open quantum system in a single experiment where the ancilla system has a
dimension equal to the total number of Kraus operators characterizing the given
quantum channel.  An arbitrary quantum channel having $d$ Kraus operators can be
simulated via DSA using unitary operations $V$, $W$, and the control operation
$U_c = \sum_{i=0}^{d-1} \vert i \rangle \langle i \vert \otimes U_i $ such that
the following condition is satisfied:
\begin{equation}
E_{k}=\sum_{i=0}^{d-1} W_{k i} V_{i 0} U_{i}  \quad (k = 0,1,2,...,d-1)
\label{dsa}
\end{equation} 
where $E_k$ is the Kraus operator, and  $V_{i0}$ and $W_{ki}$ are the elements
of $V$ and $W$, respectively. The quantum circuit for DSA is given in
Reference~\cite{xin-pra-2017}, where the initial state of the system is encoded
as $ \vert 0 \rangle_{a} \otimes \vert \psi \rangle_{s} $ which is then acted
upon by $V \otimes I$ followed by $U_c$ and $W \otimes I$, and finally a
measurement is performed on the system qubits. 

For this study, the two-qubit CBF, CPF and CBPF channels 
are characterized using
two Kraus operators as:
\begin{align}
\text{CBF} &: E_0 = \sqrt{1-p} I^{ \otimes 2}, 
\quad E_1 = \sqrt{p} \sigma_x^{ \otimes 2}  \nonumber \\
\text{CPF} &: E_0 = \sqrt{1-p} I^{ \otimes 2}, 
\quad E_1 = \sqrt{p} \sigma_z^{ \otimes 2}  \nonumber \\
\text{CBPF} &: E_0 = \sqrt{1-p} I^{ \otimes 2}, 
\quad E_1 = \sqrt{p} \sigma_y^{ \otimes 2}   
\end{align}
where $p$ is the noise strength, which can also be interpreted as probability
with which the state of the system is affected by the given noise channel. For
$p = 0$ the state of the system is unaffected, and for $p=1$ the state of the
system is maximally affected by the given noise channel. Since all the three
noise channels considered in this study have only two Kraus operators, they can
be simulated using a single ancilla qubit.  Hence for all three noise channels,
one can set $V = \left(\begin{array}{cc} \sqrt{1-p} & -\sqrt{p} \\ \sqrt{p} &
\sqrt{1-p} \end{array}\right)$, $W = I$, and $U_0  = I \otimes I$. The different
$U_1$ for CBF, CPF and CBPF channels are set to $\sigma_x \otimes \sigma_x$,
$\sigma_z \otimes \sigma_z$ and $\sigma_y \otimes \sigma_y$ respectively, such
that the condition given in Eq.~(\ref{dsa}) is satisfied.  Note that $V$ can be
interpreted as a rotation about the $y$-axis by an angle $\theta$ such that $p =
\sin^{2}{(\frac{\theta}{2})}$. 

The generalized quantum circuit using DSA to simulate all three error channels
is given in Figure~\ref{dsackt}. For the CBF channel, $U_c$ turns out to be a
Control-NOT-NOT gate, where the value of
$\theta$ (Figure~\ref{dsackt}) is zero. For the CPF and the CBPF channels, the
values of $\theta,\phi$ (the angle and axis of rotation) are $(\frac{\pi}{2},y)$
and ($\frac{\pi}{2},z$), respectively.  The output from the tomographic
measurements on the system qubits forms the column vector
$\overrightarrow{\lambda}$.  For a given value of $p$, the full vector
$\overrightarrow{\lambda}$ can be constructed by preparing the system qubits in
a complete set of linearly independent input states.

\begin{figure}[t]
\centering
\includegraphics[angle=0,scale=1]{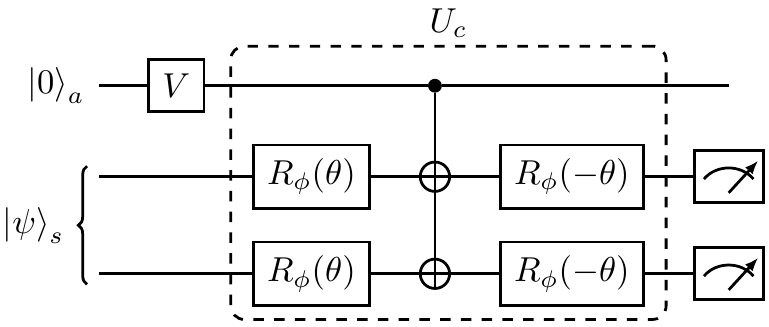} 
\caption{Quantum circuit to simulate the action of a 
correlated bit flip, a correlated phase flip  and 
a correlated bit+phase flip noise channel.
$\vert \psi \rangle_s$ are a set of linearly
independent two-qubit input states, $\vert 0 \rangle_{a}$ denotes the
state of the ancilla,
$V$ is a single-qubit rotation gate and $U_c$ denotes a set of
control operations with varying values of $(\theta,\phi)$, depending
on the noise channel being simulated.
} 
\label{dsackt}
\end{figure}
\begin{figure}[t]
\centering
\includegraphics[angle=0,scale=1.0]{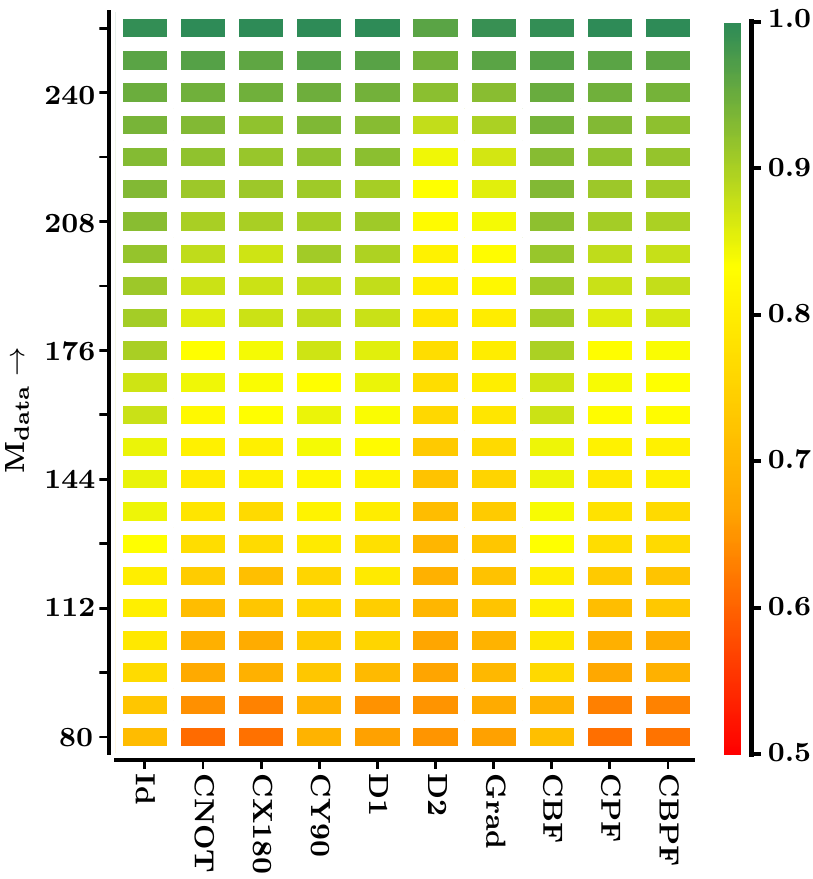} 
\caption{(Color online) Process fidelity ($\bar{\mathcal{F}}$) 
between FFNN model and the standard linear inversion method vs size
of the heavily reduced dataset
($M_{data}$), for different unitary and non-unitary quantum
processes. 
The various quantum processes are labeled on the $x$-axis and
the color coded bar on the right represents the value of the fidelity.
} 
\label{qpt}
\end{figure}

As can be seen from Figure~\ref{qpt}, the average fidelity $\mathcal{\bar{F}} =
0.8738 \pm 0.0366, 0.8272 \pm 0.0403$ and $ 0.8273 \pm 0.0416 $ for the
experimentally simulated noise channels CBF, CPF and CBPF respectively, using a
reduced data set of size 160.  Since all three correlated noise channels are
characterized by only two Kraus operators, the corresponding process matrices
turn out to be sufficiently sparse (with only two non-zero elements in the
process matrix). The FFNN can hence be used to accurately tomograph such noise
channels with arbitrary noise strength using a heavily reduced dataset.

\section{Conclusions}
\label{sec5}
Much recent research has focused on training artificial neural networks to
perform several quantum information processing tasks including tomography,
entanglement characterization and quantum gate optimization.  We designed and
applied a FFNN to perform QST and QPT on experimental NMR data, in order to
reconstruct the density and process matrices which characterize the true quantum
state and process, respectively. The FFNN is able to predict the true quantum
state and process with very high fidelity and performs in an exemplary fashion
even when the experimental data set is heavily reduced.  Compressed sensing is
another method which also uses reduced data sets to perform tomography of
quantum states and processes. However, this method requires prior knowledge such
as system noise and also requires that the basis in which the desired state
(process) is to be tomographed should be sufficiently sparse.  The FFNN, on the
other hand, does not need any such prior knowledge and works well for all types
of quantum states and processes.  Moreover, working with a heavily reduced data
set has the benefit of substantially reducing experimental complexity since
performing tomographically complete experiments grows exponentially with system
size. One can perform very few experiments and feed this minimal experimental
dataset as inputs to the FFNN, which can then reconstruct the true density or
process matrix.  Our results hence demonstrate that FFNN architectures are
promising methods for performing QST and QPT of large qubit registers and are an
attractive alternative to standard methods, since they require substantially
fewer resources.

\begin{acknowledgments}
All experiments were performed on a Bruker Avance-III 600
MHz FT-NMR spectrometer at the NMR Research Facility at
IISER Mohali. 
Arvind acknowledges funding from
the Department of Science and Technology (DST),
India, under Grant No DST/ICPS/QuST/Theme-1/2019/Q-68.
K.D. acknowledges funding from
the Department of Science and Technology (DST),
India, under Grant No DST/ICPS/QuST/Theme-2/2019/Q-74.
\end{acknowledgments}


%
\end{document}